# Sub-microsecond electric switching of large birefringence in the isotropic phase of ferroelectric nematics

*Kamal Thapa, Sathyanarayana Paladugu, and Oleg D. Lavrentovich\**

K. Thapa, S. Paladugu, O.D. Lavrentovich

Advanced Materials and Liquid Crystal Institute, Kent State University, Kent, OH 44242, USA

E-mail: olavrent@kent.edu

K. Thapa, O.D. Lavrentovich

Department of Physics, Kent State University, Kent, OH 44242, USA

O. D. Lavrentovich

Materials Science Graduate Program, Kent State University, Kent, OH 44242, USA

Conventional nonpolar nematic liquid crystals, widely known for their successful display applications, are not well suited for fast electro-optical switching because of the slow (milliseconds) field-off relaxation of the director. In this work, we demonstrate sub-microsecond field-on and field-off switching of a large (~0.1) birefringence by applying a moderate electric field $(30 - 100)$ V$\mu$m$^{-1}$, to an isotropic film of a ferroelectric nematic material. This highly efficient electro-optical switching is rooted in the recently discovered field-induced isotropic to ferroelectric nematic phase transition [J. Szydlowska *et al*., Phys. Rev. Lett. **130**, 216802 (2023)]. The demonstrated sub-microsecond switching of ferroelectric birefringence induced by field (FBIF) has the potential for applications in fast-switching electro-optic devices, such as phase modulators, light shutters, beam steerers, switchable optical compensators.

# 1. Introduction

Conventional paraelectric nematic (N) liquid crystals (LCs) are widely used in electro-optical applications enabled by anisotropy of their dielectric permittivity and birefringence.[1] Most of these applications rely on the Fréedericksz transition [2] in which an external electric field **E** reorients the apolar direction of molecular orientation called the director $\hat{\mathbf{n}} \equiv -\hat{\mathbf{n}}$, followed by a much slower relaxation when the field is switched off [1]. The only force that realigns $\hat{\mathbf{n}}$ to its original state is a relatively weak surface anchoring; the relaxation time $\tau_{\text{off}} = \frac{\gamma_1 d^2}{\pi^2 K}$ is typically in the range of milliseconds, being determined by the thickness $d$ of the cell, effective elastic constant $K$ and the rotational viscosity $\gamma_1$.[3] Numerous efforts have been explored to reduce the switching time by optimizing the dielectric and viscoelastic properties of N, using a dual frequency N, [3-4] polymer-layer-free alignment,[5] and a gradient electric field [6]. However, these methods produce only modest advancements, with the switch-off relaxation times in the millisecond to sub-millisecond range and rarely achieving 0.1 ms.[3-6]

Realignment of the director is not the only mechanism of the electro-optical response. A different approach is based on the modification of the tensorial order parameter without realignment of the director. A well-known example is the Kerr effect in which the electric field induces orientational order and birefringence.[7] Another example is the so-called electrically modified order parameter (EMOP) effect, in which the electric field changes the tensorial order parameter of the N without director realignment.[8] The EMOP and Kerr effects are fast, with switching times in the range $(1 - 100)$ ns. However, the operational fields are high, often exceeding $100 \text{ V}\mu\text{m}^{-1}$, and yielding only a modest birefringence change, $\delta n \approx 0.02$.

The discovery of ferroelectric nematic liquid crystals ($N_F$) [9] with strong spontaneous electric polarization **P** parallel to the long axes of molecules and to the director $\hat{\mathbf{n}}$ offers the opportunity to revisit the issue of electro-optic effects. The $N_F$ materials show improvements even in their N phase: a recent study of the EMOP effect demonstrated that an electric field $20 \text{ V}\mu\text{m}^{-1}$ induced $\delta n \approx 0.04$ with switching times $\approx 1$ μs. [10] The reported EMOP performance relies on the dielectric anisotropy $\Delta\varepsilon = \varepsilon_\parallel - \varepsilon_\perp$ of the material; here $\varepsilon_\parallel$ and $\varepsilon_\perp$ are the permittivities measured along $\hat{\mathbf{n}}$ and along a perpendicular direction, respectively. An intriguing question is whether a fast and robust electro-optic response can be based on the field-induced changes of the electric polarization **P**. The field-induced reorientation of molecules in the $N_F$ is triggered by a linear term $(-\mathbf{E} \cdot \mathbf{P})$ in the free energy density and thus generally occurs at lower fields than the reorientation in the N phase, controlled by the dielectric anisotropy coupling term quadratic in the field, $\left[-\frac{1}{2}\Delta\varepsilon\varepsilon_0(\mathbf{E}\cdot\hat{\mathbf{n}})^2\right]$; $\varepsilon_0$ is the electric constant. An illustration of this advantage was presented by Chen et al. for the geometry of twist Fréedericksz effect: **P** in a planar cell realigns in response to an in-plane electric field of a magnitude that is ~1000 times weaker than the operational field in the N.[9c] However, the slow switch-off time remains an issue, since it is still estimated as $\tau_{\text{off}} = \frac{\gamma_1 d^2}{\pi^2 K}$, and $\gamma_1$ in the $N_F$ phase is higher than that in the N phase.[11] Fast, on the order of 1 µs, oscillatory reorientation of **P** can be induced in a planar $N_F$ cell by an out-of-plane alternating current (ac) electric field, [12] which is similar to the splay Fréedericksz effect in the N. However, the angular amplitude of oscillatory realignment is weak, less than 1°, which does not produce a large change of birefringence.[12]

In contrast to the Fréedericksz-type realignment of **P**, very little is known about the electro-optical response of the $N_F$ in the Kerr and EMOP settings, when the polar order is either induced or modified by the electric field. Recent research by Szydlowska *et al*. demonstrates a Kerr-like effect in which the electric field induces polar ferroelectric ordering in the isotropic phase of an $N_F$ material. [13] However, it is not known how fast this response is.

In this work, we explore the dynamics of the field-induced polar ordering of the isotropic phase of two $N_F$ materials and demonstrate that a modest electric field of an amplitude $(30 - 100)$ Vµm$^{-1}$ induces a strong birefringence $\delta n \approx 0.1$ with response times $\approx (0.1 - 1.0)$ µs in both the field-on and field-off cycles. The ferroelectric birefringence induced by field (FBIF) shows a sigmoid dependence on the field amplitude, with a critical slowing down for the intermediate amplitudes of the field. The figure of merit of the field-induced birefringence caused by the polar ordering of molecules is significantly higher than the figure of merit of the Fréedericksz and EMOP effects in the paraelectric N. The work demonstrates the potential of ferroelectric nematic materials in fast electro-optical devices.

## 2. Results

We explore two room-temperature $N_F$ materials, namely, UUQU-4-N [14] and FNLC919 (both received as a gift from Merck KGaA, Darmstadt, Germany). The phase sequence of UUQU-4-N on cooling is I – 20 °C - $N_F$, [14] and of FNLC919 is I – 82 °C – N - 46 °C – $N_x$ – 32 °C - $N_F$ – 8 °C – Cr [15]. Here $N_x$ stands for an apolar phase and Cr stands for a crystal. The materials are placed in a sandwich cell of a thickness $d \approx 5.4$ µm. The out-of-plane field-induced dynamic birefringence is tested by a laser beam of a wavelength 632.8 nm impinging obliquely at the cell, following the procedure described previously, [10] **Figure 1**.

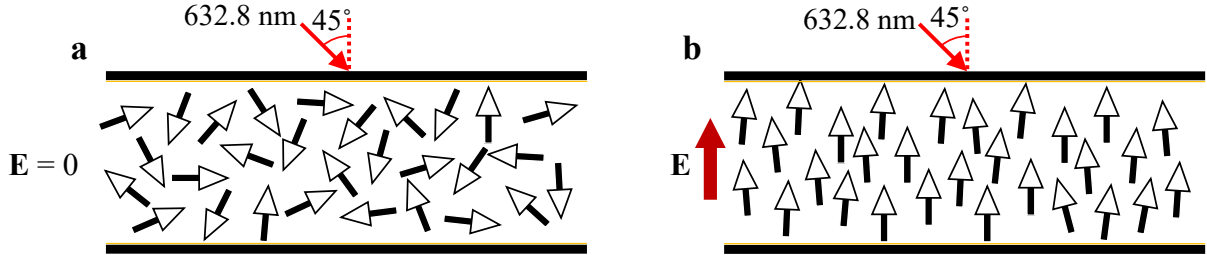

**Figure 1.** Ferroelectric birefringence induced by the field (FBIF): a) Isotropic phase of zero birefringence in the absence of the electric field; b) birefringence caused by ferroelectric nematic orientational order in an applied electric field **E**.

## 2.1. Ferroelectric nematic material UUQU-4-N

Applying a short volage pulse of duration 2 μs, with sharp front and end edges (less than 10 ns), to the cell filled with the isotropic phase of UUQU-4-N induces a strong time-dependent birefringence $\delta n(t)$, which reaches its maximum value $\delta n_{max} \approx 0.12$ at relatively moderate electric fields $\approx (70 - 100)$ Vμm$^{-1}$, **Figure 2**. For example, a pulse of 94 Vμm$^{-1}$ induces birefringence $\delta n$ that grows from zero to $\delta n_{max} \approx 0.11$ within $\tau_{on} = 80$ ns and then returns to zero within $\tau_{off} = 60$ ns. Therefore, the observed FBIF shows a remarkably fast submicron switching times in both field-on and field-of regimes. The birefringence changes in Figure 2a are achieved at a temperature of 100 °C, which is 80 °C higher than the temperature of the I-to-N$_F$ phase transition. The higher the voltage, the higher the induced birefringence, Figure 2b. The maximum field-induced birefringence $\delta n_{max}$ as a function of the 2 μs electric pulse amplitude is of a sigmoid shape, Figure 2c. The behavior of $\delta n_{max}(E, T)$ in Figure 2c is similar to the field-temperature curves of retardance (birefringence times the cell thickness) measured in a different N$_F$ material and attributed to the field-induced phase transition from the isotropic phase to the phase with polar ordering of molecules.[13]

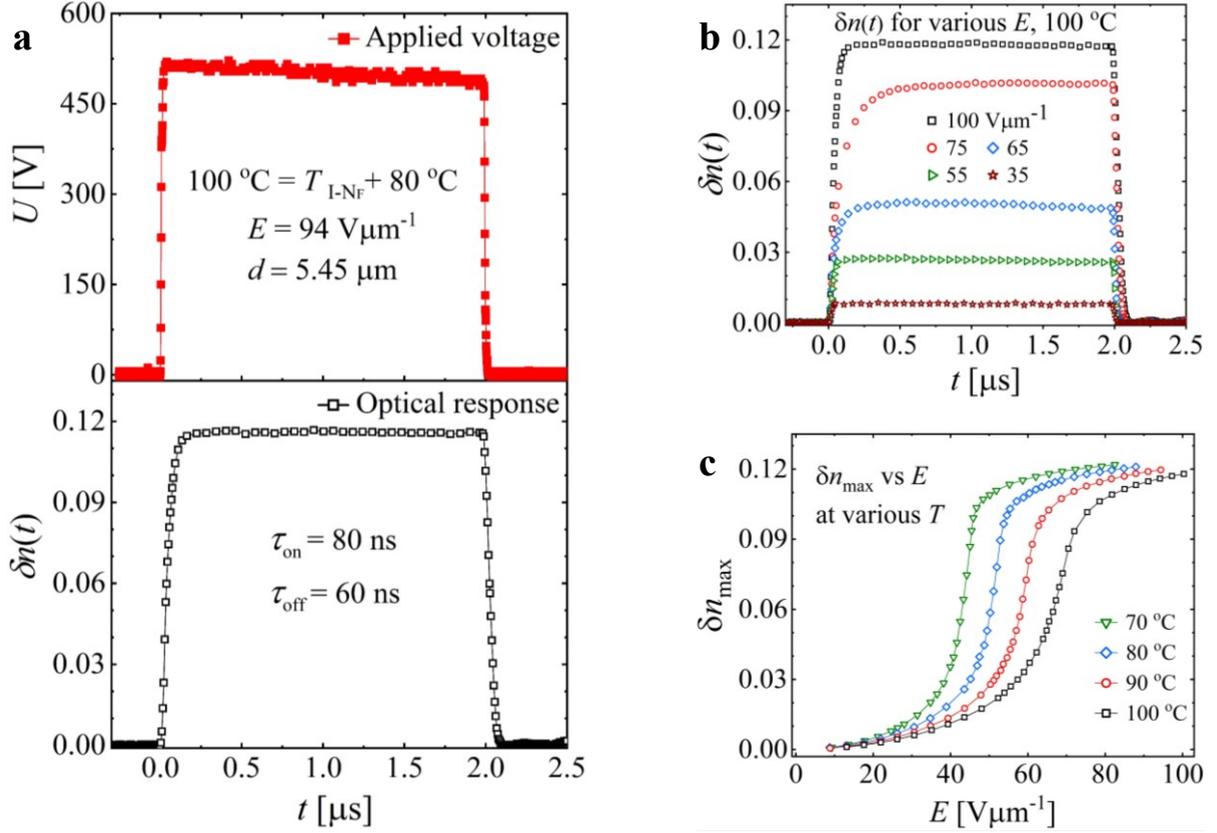

**Figure 2.** Electric field dependency of optic response $\delta n(t)$ in the isotropic phase of UUQU-4-N for a cell of thickness $d = 5.45$ μm: a) Voltage pulse of duration 2 μs and amplitude 510 V causes rapid increase of birefringence $\delta n(t)$ at 100 °C; b) $\delta n(t)$ at 100 °C for different electric pulse amplitudes; c) electric field dependence of $\delta n_{max}$ at various temperatures.

Time-resolved measurements of dynamics reveal an important feature of the field-on switching time $\tau_{on}$, namely, the existence of critical slowing down for intermediate electric fields, **Figure 3a**. These intermediate fields correspond to the middle of the sigmoid $\delta n_{max}(E)$ dependencies in which the derivative $\partial(\delta n_{max})/\partial E$ is maximum, Figure 2c. In these regions, $\tau_{on}(E)$ increases sharply, reaching a maximum at some $E_{csd}$ and then regains its shorter values when $E > E_{csd}$, Figure 3a; the subscript "csd" stands for critical slowing down. The qualitative explanation is that near $E_{csd}$, the energy minima of the two states, a paraelectric with ferroelectric fluctuations, and a ferroelectric with the field-induced saturation of the degree of orientational order, are close to each other with a vanishing energy barrier, which is the common mechanism of critical slowing down in many transitions, see, for example, a review.[16] In contrast, $\tau_{off}$ shows a continuous increase with the field at $E > E_{csd}$, Figure 3b. A tentative explanation is that the field-induced ferroelectric phase remains in a metastable state separated from the isotropic state by a high energy barrier. Both $\tau_{on}$ and $\tau_{off}$ become longer as the temperature drops below 70 °C; for example, at 50 °C, $\tau_{on}(E = E_{csd}) = 2.2$ µs while $\tau_{off}(E = E_{csd}) = 0.5$ µs; the increase is associated with the higher viscosity.

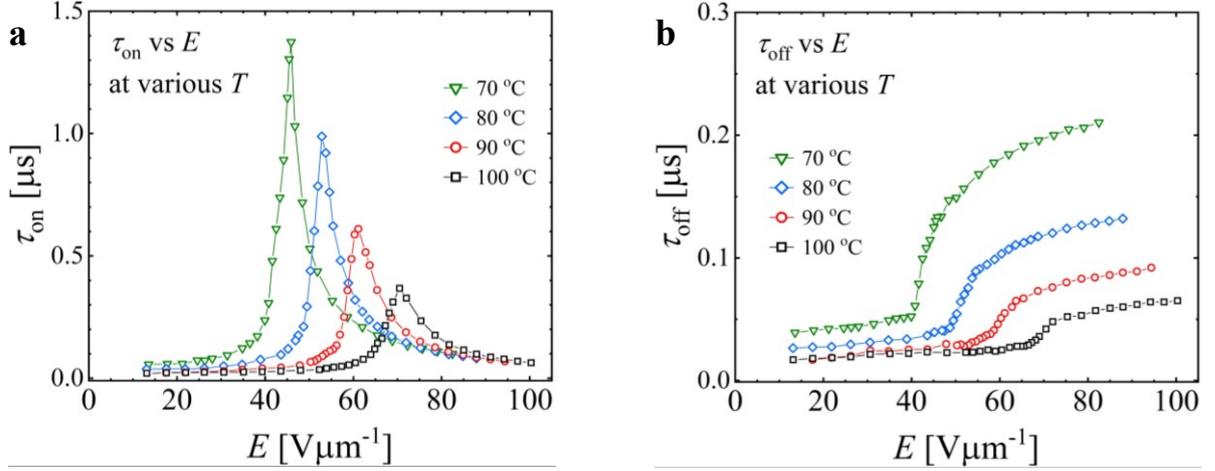

**Figure 3**. Electric field dependence of a) switch-on ($\tau_{on}$) and b) switch-off ($\tau_{off}$) times in the isotropic phase of UUQU-4-N at various temperatures. The pulse duration is 2 µs.

**2.2 Ferroelectric nematic mixture FNLC919**

To verify whether the reported dynamics of electro-optical response is a general feature of the $N_F$ materials, we explore the mixture FNLC919. The FBIF dynamics in FNLC919 is similar to that in UUQU-4-N but is shifted to somewhat higher temperatures, lower electric fields, and longer response times. As shown in **Figure 4a**, a 215 V pulse with a duration of 3 µs at 100 °C induces $\delta n_{max} \approx 0.1$ with $\tau_{on} \approx \tau_{off} = 0.2$ µs. Although slower than UUQU-4-N, FNLC919 operates at a nearly 2.5 times lower driving field. The maximum field-induced birefringence $\delta n_{max}$ increases with the field, reaching a saturation value $\approx 0.11$, Figure 4b,c.

The switching times $\tau_{on}$ and $\tau_{off}$ of FNLC919 exhibit similar field dependences as their UUQU-4-N counterparts, **Figure 5**.

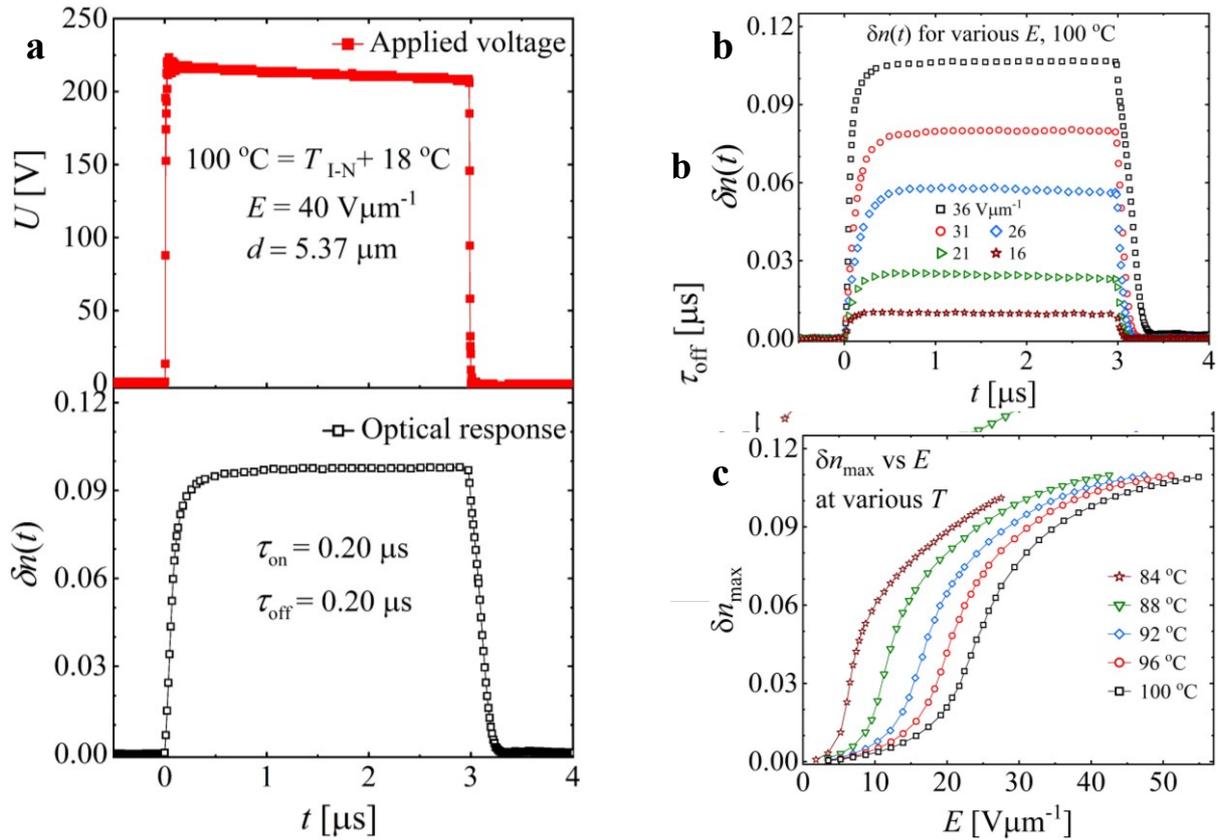

**Figure 4.** Electric field and temperature dependencies of optic response $\delta n(t)$ in the isotropic phase of FNLC919 for a cell with thickness = 5.37 μm: a) $\delta n(t)$ at 100 °C under a 215V pulse; b) $\delta n(t)$ at 100 °C with varying electric fields; c) electric field dependence of $\delta n_{max}$ at various temperatures. The pulse duration is 3 μs.

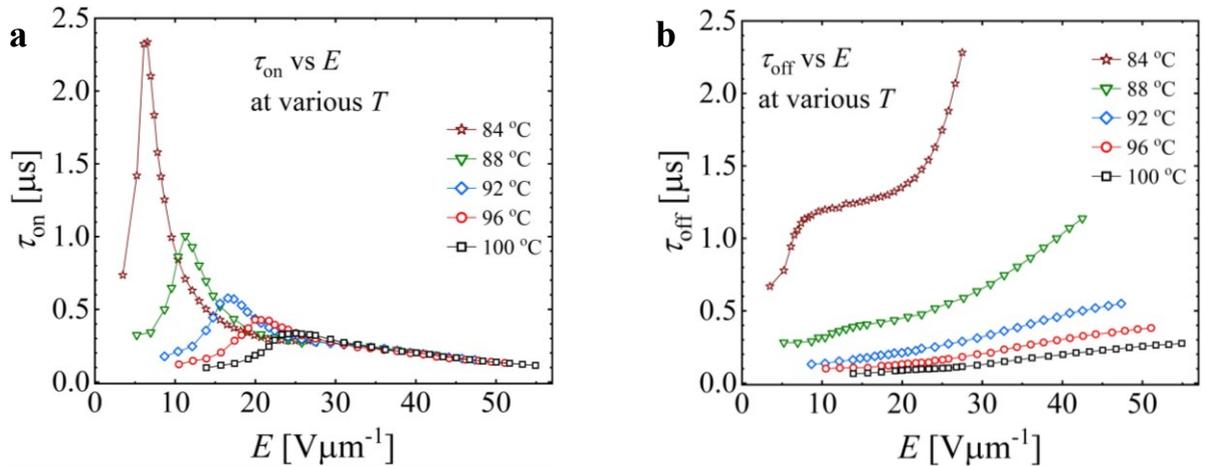

**Figure 5**. Electric field dependence of a) switch-on ($\tau_{on}$) and b) switch-off ($\tau_{off}$) times in the isotropic phase of FNLC919 at various temperatures. The pulse duration is 3 μs.

## 3. Discussion

In electrically controlled LC-based optical retarders, the primary objective is to achieve a sufficiently large optical retardance, $\delta\Gamma_{\max} = \delta n_{\max} \times d$, of at least half the probing wavelength ($\lambda/2$) in the shortest possible time. The performance is quantified by a figure of merit (FoM) defined as $\text{FoM} = \delta\Gamma_{\max}^2/\pi^2\tau_{\text{off}}$, which depends on $\tau_{\text{off}}$ [3, 17]. We chose the equality $\tau_{\text{off}} \approx \tau_{\text{on}}$ at $E > E_{csd}$ as the criterium for the evaluation of FoM of FBIF. At 100 °C, FBIF in UUQU-4-N exhibits FoM=$5.7 \times 10^5$ $\mu m^2 s^{-1}$, nearly 200 times higher than the FoM of the EMOP in the nematic phase of $N_F$ material. [10] **Table 1** compares FoMs across different modes of LC electro-optic switching.

**Table 1**. Comparison of the electro-optical performance of different LC switching modes that do not involve director realignment; NEMOP stands for nanosecond electric modification of order parameter; MEMOP stands for microsecond electric modification of order parameter, FBIF stands for the ferroelectric birefringence induced by the field.

| Material type | Method | Material | T [°C] | E [V$\mu m^{-1}$] | $\delta n_{\max}$ | $\tau_{\text{on}}$ | $\tau_{\text{off}}$ | FoM[a] [$\mu m^2 s^{-1}$] |
|---|---|---|---|---|---|---|---|---|
| Paraelectric nematic LCs | EMOP [8c] | HNG715600-100 | 20 | 170 | 0.013 | 30 ns | 30 ns | $1.4 \times 10^4$ |
| | Enhanced EMOP [7h] | HNG715600-100 doped with DPP (19 wt.%) | 23 | 190 | 0.020 | 130 ns | 70 ns | $1.4 \times 10^4$ |
| | Kerr [7g] | 5CB | 75 | 115 | 0.035 | 10 ns | 5 ns | $6.2 \times 10^5$ |
| | EMOP [18] | GPDA200 | 72 | 28 | 0.038 | 7 μs | 13 μs | $2.8 \times 10^2$ |
| Ferroelectric nematic LCs | EMOP: Nematic phase [10] | FNLC919 | 80 | 21 | 0.040 | 0.65 μs | 1.3 μs | $3.0 \times 10^3$ |
| | FBIF: Isotropic phase [This work] | UUQU-4-N | 100 | 94 | 0.116 | 80 ns | 60 ns | $5.7 \times 10^5$ |
| | | | 90 | 88 | 0.118 | 90 ns | 90 ns | $3.9 \times 10^5$ |
| | | | 80 | 75 | 0.118 | 0.12 μs | 0.12 μs | $2.9 \times 10^5$ |
| | | | 70 | 65 | 0.118 | 0.18 μs | 0.19 μs | $1.9 \times 10^5$ |
| | | FNLC919 | 100 | 40 | 0.098 | 0.20 μs | 0.20 μs | $1.2 \times 10^5$ |
| | | | 96 | 33 | 0.093 | 0.24 μs | 0.23 μs | $0.95 \times 10^5$ |
| | | | 92 | 28 | 0.088 | 0.28 μs | 0.29 μs | $0.68 \times 10^5$ |
| | | | 88 | 17 | 0.072 | 0.43 μs | 0.42 μs | $0.31 \times 10^5$ |

a) FoMs are calculated for the cell thickness $d = 5$ μm

The FoMs of FBIF achieved in this work are on the order of $10^5$ μm$^2$s$^{-1}$, which are at least three to four orders of magnitude higher than the FoM of the conventional Fréedericksz effect, which shows the range $(1 - 10)$ μm$^2$s$^{-1}$ [3, 17, 19]. These values are also comparable to or higher than the FoMs of EMOP [7h, 8a-c, 10, 18] effects in paraelectric N. The only exception is the Kerr effect described for the paraelectric nematic pentylcyanobiphenyl (5CB) [7g], in which the FoM is high mostly because of the very short $\tau_{\text{off}} = 5$ ns; however, the field induced birefringence in 5CB is three times weaker than the birefringence reported in our work for the N$_F$ materials. Since the FBIF reaches $\delta n_{\max} \sim 0.1$ at the electric fields below 100 Vμm$^{-1}$, it allows one to produce an optical retardance change $\delta \Gamma \approx 500$ nm, in a thin ($\approx 5$ μm) N$_F$ cells; this retardance is well above the $\lambda/2$ requirement for the entire visible range and beyond.

## 4. Conclusion

We demonstrate that the dynamics of the electrically induced polar ordering in the isotropic phase of two ferroelectric nematic materials is extremely fast and robust, causing a ferroelectric birefringence induced by field (FBIF) on the order of 0.1 that raises when the field is on and decays when the field is off within about 100 ns. In both materials, the induced birefringence and switching times show strong dependence on the electric field amplitude and temperature. In general, the switching times decrease as the temperature increases, thanks to a lower viscosity. Intermediate values of the electric field cause a critical slowdown of the response to field-on switching; however, the response becomes fast again as the electric field is increased past the intermediate values. The induced levels of birefringence, 0.1, are extraordinarily high, being at least three times higher than the best results achieved in the paraelectric nematic materials, which we associate with the high degree of orientational order in the field-induced state.

It is important to note that the operational fields reported above did not cause an electric breakdown of the cells. When the voltage applied to the cells of thickness $d \approx 5.4$ μm were elevated above 600 V in the case of UUQU-4-N and 400 V in the case of FNLC919, some cells could burn; however, if a cell survived initial application of these elevated voltages, it would remain functional for repeated experiments.

This FBIF phenomenon represents a new prominent member of the broad class of fast electro-optic switching mechanisms that embraces Kerr, [7, 20] and EMOP [7h, 8, 10, 18] effects. The demonstrated performance of N$_F$ materials shows their potential for applications in fast-switching electro-optical devices such as phase modulators, light shutters, beam steerers, switchable optical compensators, etc. It is expected that the exploration of N$_F$ materials other than UUQU-4-N and FNLC919 would result in even better performance. A theoretical description of the FBIF effect, dynamics of field-induced paraelectric and ferroelectric states, would optimize the search; this work is in progress.

## 4. Experimental Section

The molecular structure of $N_F$ material UUQU-4-N is shown in **Figure 6.**

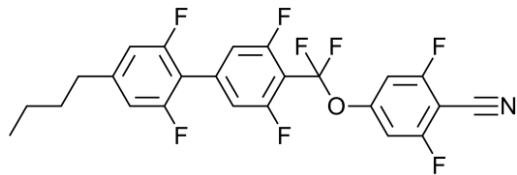

**Figure 6.** Molecular structure of UUQU-4-N.

*Polarization Measurements:* The spontaneous polarization in the $N_F$ phase of the materials is measured at 3 °C below the phase transition temperature to the $N_F$ phase, using planar (parallel assembly) cells of a thickness ≈ 15 μm. Each cell is assembled using two glass plates: one is a plain glass plate (without ITO) coated with the polyimide PI2555 (HD Microsystems LLC) and buffed, while the other plate has the buffed PI2555 over two in-plane ITO electrodes separated by a 1 mm gap. To measure the spontaneous polarization, a triangular waveform of frequency 50 Hz with 175 V peak-to-peak (waveform generator: SDS 1032X SIGLENT and wideband voltage amplifier: 7602 KROHN-HITE Corp.) is applied across a series circuit of the LC cell and 20 kΩ resistor. The voltage drop across the series resistor is recorded using a digital oscilloscope (TBS2000B, Tektronix). Then the polarization current, polarization charge (time integral of polarization current), and the spontaneous polarization are calculated as described in ref.[21]

The measured polarization densities in the $N_F$ phase are $7.6 \times 10^{-2}$ $Cm^{-2}$ in UUQU-4-N, and $5.4 \times 10^{-2}$ $Cm^{-2}$ in FNLC919, **Figure 7**. These values are comparable to those of other $N_F$ materials such as DIO ($\approx 5 \times 10^{-2}$ $Cm^{-2}$) [21] and RM734 ($\approx 6 \times 10^{-2}$ $Cm^{-2}$). [9c]

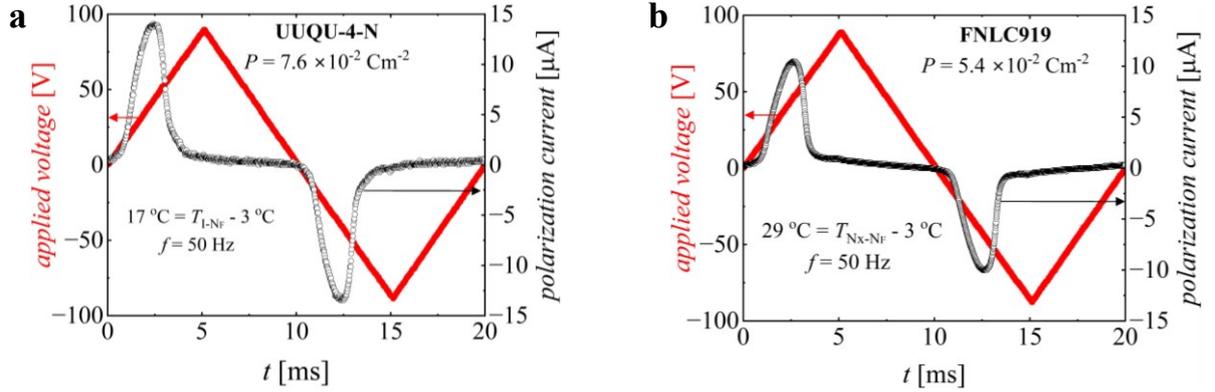

**Figure 7.** Time dependence of polarization current under a triangular pulse for a) UUQU-4-N and b) FNLC919 in the $N_F$ phase, at a common reference temperature 3 °C below the phase transition temperature to the $N_F$ phase.

*Electro-optical Measurement:* The electro-optical response of the materials in their isotropic phases is studied in the temperature ranges $(70 - 100)$ °C for UUQU-4-N and $(84 - 100)$ °C for FNLC919, using flat planar cells. Each cell is made up of two glass plates with patterned ITO of a small area of 2 mm × 2 mm and low sheet resistance of 10 Ω/□. Both glass plates are coated with the polyimide SE5661 (Nissan Chemicals, Ltd.). The plates are separated by a distance $d \approx 5.4$ µm. A He-Ne laser beam ($\lambda = 632.8$ nm) passes through two crossed linear polarizers with the cell and a Soleil-Babinet compensator positioned between them, **Figure 8a**. The probing laser beam is incident at 45° to the normal to cell, Figure 1. The cell is sandwiched between two right-angled prisms and is placed inside a custom-made copper holder. This entire assembly is further placed in a Teflon holder for thermal insulation. The transmitted light intensity is measured using a photodetector TIA-525 (Terahertz Technologies, response time < 1 ns). Voltage pulses with sharp rise and fall edges are applied to the LC cell by a system of waveform generator WFG500 (FLC Electronics), high DC voltage source KEITHLEY 237 (Keithley), and pulse generator HV 1000 pulser (Direct Energy). The input voltage pulses, and the output photodetector signals, Figure 8b, c, are monitored using a digital oscilloscope (TBS2000B, Tektronix). The temperature of the sandwiched cell is controlled by a hot stage LTS120 and a controller PE94 (both Linkam), with an accuracy of 0.01°C.

The dynamics of the field-induced birefringence, $\delta n(t)$, is characterized by a four-point measurement scheme, described in Ref. [18]. The technique uses four successive measurements of the transmitted intensities $I_k$, $k = 1, 2, 3, 4$, for the same area of cell subjected to identical electrical pulses at four different settings of the retardance $\Gamma_k^{SB}$ of Soleil-Babinet compensator,[22]

$$I_k(t) = [I_{\max}(t) - I_{\min}(t)] \sin^2\left[\frac{\pi}{\lambda}\Gamma_k(t)\right] + I_{\min}(t), \qquad (1)$$

where $\Gamma_k(t) = \Gamma_N(0) + \Gamma_k^{SB} + \delta\Gamma(t)$ is the total dynamic optical retardance of the system and $\Gamma_N(0)$ is the optical retardance of the LC cell in the isotropic phase when there is no electric field. We select $\Gamma_k^{SB} = \lambda(m + k/4) - \Gamma_N(0)$, where $m$ is an integer, in such a way that the initial intensities with zero field are maximum $I_2(0)$, minimum $I_4(0)$, and mean values $I_1(0) = I_3(0) = [I_2(0) + I_4(0)]/2$. For this selection, the field-induced birefringence is calculated as

$$\delta n(t) = \frac{\lambda}{2\pi d} \arg\{I_2(t) - I_4(t) + i[I_1(t) - I_3(t)]\}. \qquad (2)$$

As an example, the dynamics of transmitted light intensities $I_1(t)$ and $I_3(t)$ are presented in Figure 8b, while $I_2(t)$ and $I_4(t)$ are shown in Figure 8c, under a voltage pulse of duration 2 µs with an electric field amplitude of 94 Vµm$^{-1}$ at 100 °C, for UUQU-4-N. The switching-on ($\tau_{\mathrm{on}}$) and switching-off ($\tau_{\mathrm{off}}$) times are defined as the durations within which $\delta n(t)$ changes between 10% and 90% of its maximum value, $\delta n_{\max}$.

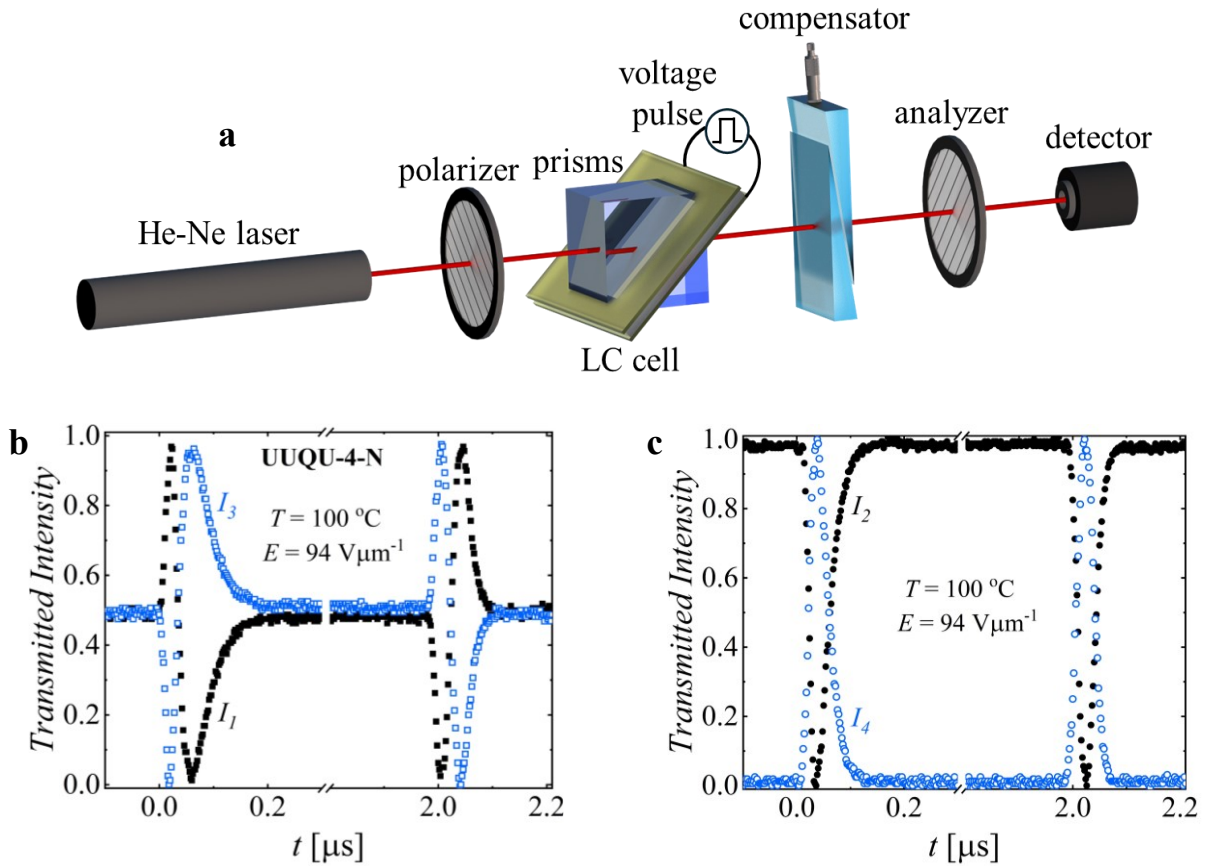

**Figure 8.** a) Experimental setup: an LC cell is sandwiched between two right angle prisms, an impingent He-Ne laser beam makes an angle 45° with the normal to cell and the direction of the applied electric field. Dynamics of normalized transmitted light intensities: b) $I_1$ and $I_3$, c) $I_2$ and $I_4$ at four different compensator's settings. The pulse duration is 2 µs.

**Acknowledgments**

We are grateful to S.V. Shiyanovskii for fruitful discussions. We thank M. Klasen-Memmer and R. Tuffin, Merck KGaA, Darmstadt, Germany for the kindly provided materials UUQU-4-N and FNLC919. The work was supported by NSF grant DMR-2341830.

**Conflict of Interest**

The authors declare no conflicts of interest.

**Data Availability Statement**

The Data that supports the finding of this study are available from the corresponding author upon reasonable request.